\documentstyle[11pt,newpasp,twoside,epsf]{article}
\def\edcomment#1{\iffalse\marginpar{\raggedright\sl#1\/}\else\relax\fi}
\reversemarginpar

\begin{document}
\title{Dissecting the luminosity function of the Coma cluster of galaxies
by means of CFHT wide field images}
\author{S. Andreon} 
\affil{Osservatorio Astronomico di Capodimonte, via Moiariello 16, 80131
Napoli, Italy}
\author{J.-C. Cuillandre}
\affil{Canada--France--Hawaii Telescope, PO Box 1597, Kamuela, 96743, Hawaii}

\begin{abstract} By means of wide--field ($42\times28$ arcmin) CCD images of
the Coma cluster of galaxies and of control fields, taken at the
Canada--France--Hawaii telescope with the CFH12K and UH8K cameras, we
determine the luminosity function bi--variate in central brightness. We found
a clear progression for a steeping and weakening of the FL going from high
surface brightness galaxies ($\mu\sim20$ mag arcsec$^{-2}$) to galaxies of
very--faint central brightnesses ($\mu\sim24.5$ mag arcsec$^{-2}$). Compact
galaxies, usually rejected in the star/galaxy classification, 
are found to be a minor
population in Coma. A huge population of low surface brightness galaxies red
and faint are discovered, representing the largest contributor to the LF at
faint magnitudes. Among them, there could be the remnant of high redshift
starbursting galaxies responsible for the Butcher--Oemler effect. 
\end{abstract}

\section{Introduction}

The luminosity function (LF hereafter), i.e. the number density of galaxies having
a given luminosity, is critical to many observational and theoretical problems (see
e.g. Binggeli, Sandage \& Tammann 1988). From an observational point of view, the
LF is the natural ``weight" of all those quantities which need to be weighted
against the relative number of objects in each luminosity bin. Furthermore, due to
the role played by luminosity and surface brightness in the inclusion of objects in
any observed sample (faint objects or low surface brightness galaxies are often
excluded or under--represented), the knowledge of the LF and the LF bi--variate in
brightness is fundamental to compute the selection function and is needed to derive
the actual galaxy properties from the measured quantities (see, for example, the
discussion on the field LF steepness by Sprayberry et al. 1997). 

The LF buries under its foots the true problem (Sandage 1994): the LF is likely the
sum of the LFs of the specific types, or of any other physically based galaxy
classes. Galaxies can be classified on the bases of their central
brightness, and in fact depending on their central brightness they lay in different
parts of the Fundamental Plane (e.g. Bender, Burstein \& Faber 1997), showing that
this classification is not an aestheatical one but reflects some physical difference
between the classes. Therefore, ``it would be of great importance to know what the
luminosity function looks like when divided into classes of surface brightness"
(Kron 1994).

In the present paper we shortly present the LF bi--variate in surface brightness of
a sample of galaxies in the Coma cluster of more than 4000 members and in three
filters, down to the magnitude of three bright globular clusters, complete in
surface brightness down to the brightness of the faintest cataloged LSB galaxies,
spread up over the largest cluster area ever observed with CCDs. The interested
reader may found details in Andreon \& Cuillandre (2000).

\section{The CFHT data}
 
$B$, $V$ and $R$ Coma cluster observations have been taken during the CFH12K 
(Cuillandre et al. 2000, see also Cuillandre et al. in this proceeding) first 
light at the Canada--France--Hawaii telescope prime focus in photometric 
conditions. CFH12K, the third generation wide-field imager at CFHT, is a 
12K $\times$ 8K backside-illuminated CCD mosaic 
camera, offering a field of view of 42 $\times$ 28 arcmin$^2$ with a pixel 
size of 0.206 arcsec. 
A total of 12\,mn were accumulated in each filter, split in four dithered 
3\,mn exposures to remove blemishes and cosmic rays and most importantly 
to fill the 6 arcseconds gaps between the CCDs. The data were reduced
using standard CCD data reduction procedures (additive and multiplicative
components, alignment and combination) implemented in a software specially 
optimized for fast processing of large CCD mosaics data at CFHT
(FLIPS, Cuillandre 2001).

The control fields have been observed with the same telescope, with the
same camera ($B$) during the same night, and using the UH8K ($V$ and $R$)
in 1998.  UH8K, a 8K $\times$ 8K frontside-illuminated CCD mosaic camera
was the second generation wide-field imager at CFHT. The $B$ data is a
deep (2\, hours integration) empty field nearby the galaxy NGC\,3486, a
field standing at a similar galactic latitude as the Coma cluster. The $V$
and $R$ (3\,hours and 2\,hours resp. of integration) data are centered on
the SA\,57 field, also at a similar galactic latitude to Coma. The control
fields were reduced in an indentical way as the Coma cluster data, and
then the properties (noise and PSF) of these images are matched to those
of Coma.

\begin{figure}
\plotfiddle{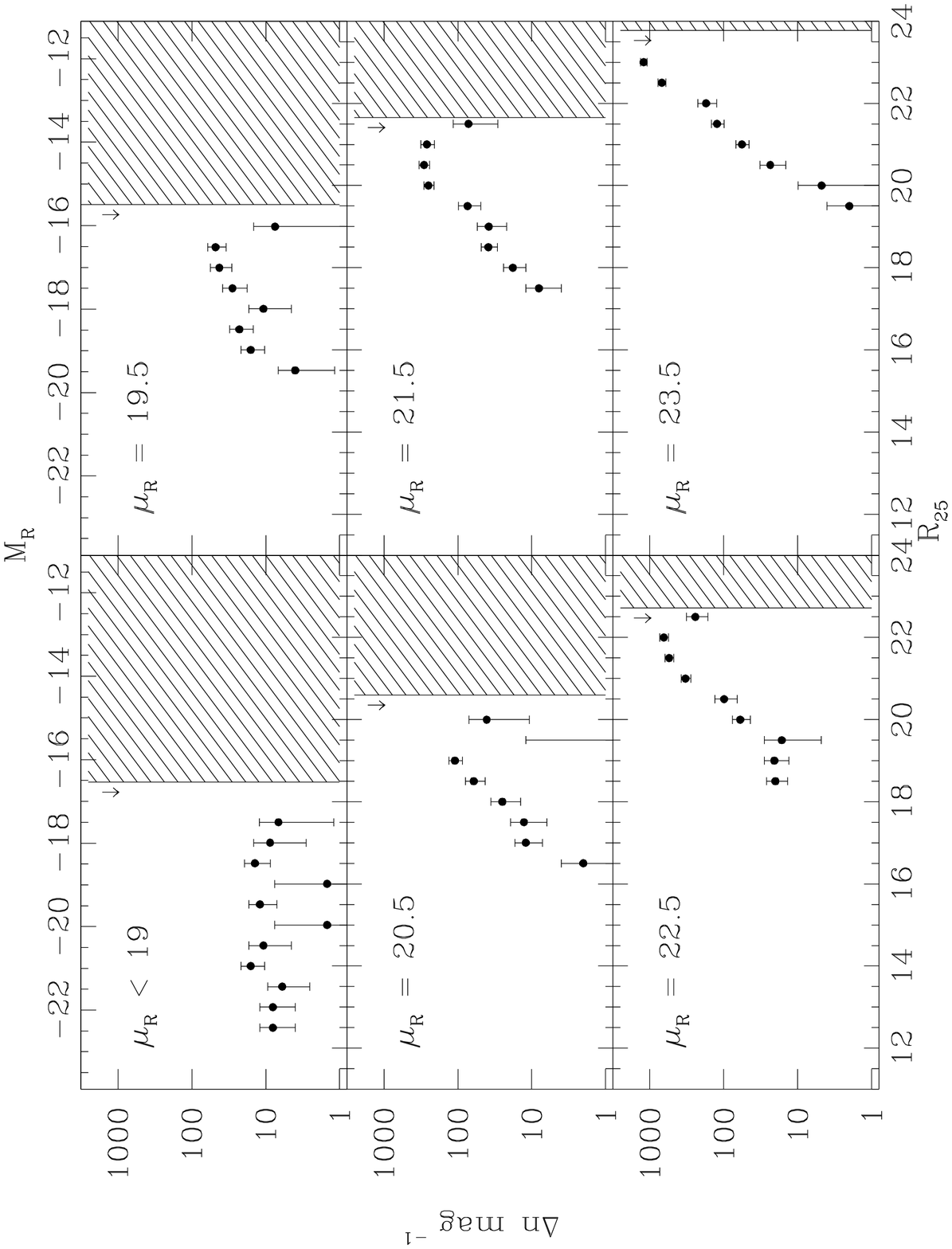}{9truecm}{-90}{50}{50}{-200}{270}
\plotfiddle{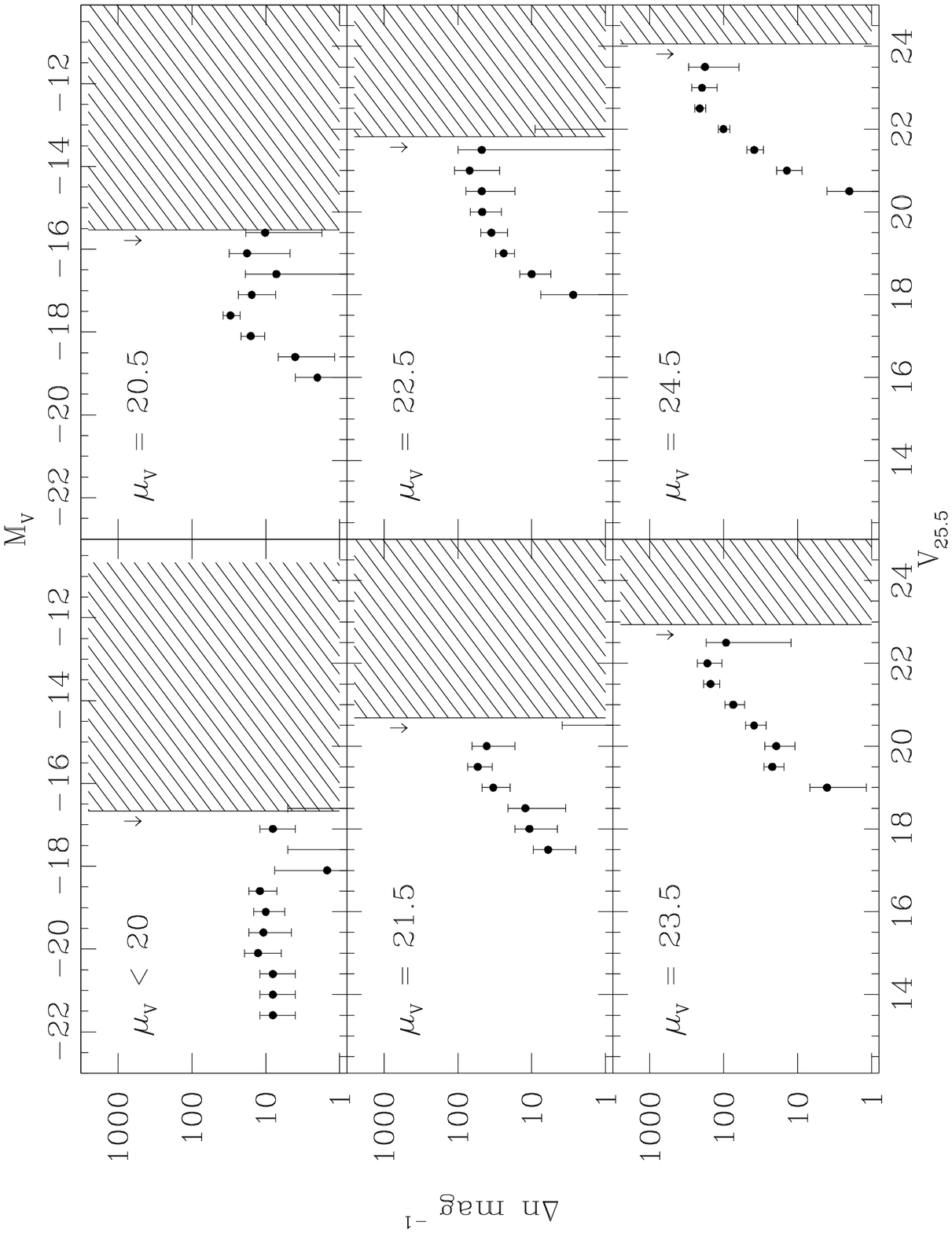}{9truecm}{-90}{50}{50}{-200}{270}
\caption{Bi--variate LF of Coma galaxies in the $R$ and $V$ bands. 
There is a clear progression from flat and bright LFs of HSB galaxies 
to steep and faint LFs of LSB. Errorbars are as in Figure 3.}
\end{figure}

\section{The bi--variate luminosity function}

The bi--variate LF is computed as the statistical difference between
counts in the Coma and control field directions, in order to remove the
background contribution from counts in the Coma direction. Figure 1 shows
the $R$ and $V$ bi--variate LF. This is the first so far accurately
computed for any environment, to our best knowledge. Brightness bins are 1
mag arcsec$^{-2}$ wide. The previous larger effort in this direction is
presented in de Jong (1996), which study a sample of 86 field galaxies,
while our sample includes $\sim 3000$ cluster member galaxies. Central
brightness is the actual galaxy central brightness convolved with the
seeing disk (whose FWHM correspond to $\sim1$ kpc to the Coma cluster
distance) and then measured in a 0.25 kpc aperture.

At all brightness bins, galaxies occupy a bounded range. Although a
distribution with a finite width is expected, we can now quantify it for
the first time. Galaxies of very large size or very flat surface
brightness profile (i.e. near the left end of each bi--variate LF plot)
are uncommon, with a relative frequency distribution presented in Figure
1.

At the bright end, galaxies more than 4 mag brighter than their central
brightness are very uncommon.  Bi--variate LFs are bounded at the faint
end too, because any object can be fainter (and smaller) than an
unresolved point source (star) of the same central (apparent) brightness.
An arrow in the plots marks this magnitude. Seldom the point at the
pointed magnitude, i.e. objects that are as compact as the seeing disk, is
on the extrapolation of points at brighter magnitudes. The rarity of
galaxies at the arrow magnitude, when compared to the expected location
based on the trend at brighter magnitudes, is not due to the fact that
compact galaxies are removed in the star/galaxy classification or
implicitly supposed not to exist, because our derivation of the LF does
not follow this path: at the difference of most previous works, we have
not removed compact dwarfs in the star/galaxy classification or implicitly
supposed that they do not exist. The found rarity of compact galaxies
means that most of the galaxies at the Coma distance are extended sources
at our resolution. 

Lacking a {\it field} bi--variate LF, it is difficult to say whether the
Coma cluster is effective in harassing LSB galaxies, as advocated by Moore
et al. (1996; 1999) or the bi--variate LF is the same in the two
environments and it tells us more on galaxy formation and evolution in
general.

The comparison of Figure 1 panels shows that LSB dominate the LF at faint
magnitudes, while high surface brightness (HSB) galaxies dominate the
bright end.  High surface brightness galaxies ($\mu_0<20.0$ mag
arcsec$^{-2}$)  have a shallow LF ($\alpha\sim-1$), while LSB galaxies
have a steep and fainter LF. There is a clear trend for a steepening and
weakening of the FL going from high surface brightness galaxies to faint
and very faint central brightnesses. 

From the comparison of the Coma LF at different wavelenghts, we found
that faint Coma galaxies are red (Andreon \& Cuillandre 2000; Andreon,
Cuillandre \& Pell\'o 2000). Faint galaxies are mostly of low surface
brightness (LSB), and therefore in Coma faint galaxies are red and of LSB.
In the field, LSB galaxies dominate the LF at faint magnitudes (Sprayberry
et al. 1997) and are preferentially blue (de Blok et al. 95; van der Hoek
et al. 2000). However, red LSB have been recently discovered in a survey
targeting mainly two nearby clusters (O'Neil, Bothun \& Cornell 1997), but
the large majority are still blue. Therefore LSBs in Coma and in the field
have quite different colors.

The existence of a large population of red LSB galaxies in the Coma
cluster has a significant impact in the context of the Butcher--Oemler
(1978, 1984) effect. It has been argued by Rakos \& Schombert (1995), and
then by many other authors later, that unobserved descendents of the large
population of starbursting high redshift galaxies represent a problem for
the Butcher--Oemler effect. LSB galaxies has been excluded as descendent
because they have not the properties of failed galaxies, lacking, for
example, the expected correlation between color and brightness for a
fading population (Bothun, Impey \& McGaugh 1997). However, the LSB
galaxies disclaimed to be failed galaxies are blue and in the field while
exhausted galaxies should be red and in cluster, as Coma LSB galaxies are.
Therefore, we suggest a too premature dismiss of the LSB galaxies as a
remnant of high redshift starburst galaxies. There is ample room among
several thousand red LSBs in the Coma cluster for accounting for a few
tens of faded remnants and therefore Coma could be, from this point of
view, the descendent of clusters with high blue fractions. In this
context, we are assuming that the Butcher--Oemler effect is still a
measure of the galaxy evolution, even if it could be largely plagued by
selection effects, as shown by Andreon \& Ettori (1999).

{\bf Acknoledgements}

The authors wish to thank P. Couturier, director of CFHT, for the
allocation of discretionnary time for this program. The authors thank M.
Cr\'ez\'e and A. Robin for sharing their UH8K SA\,57 field data. S. A.
warmly thanks the CFHT corporation for financial support to attend this
conference.

\end{document}